\documentstyle[floats,aps,psfig,pre]{revtex}
\begin{document}
\flushbottom
\twocolumn[\hsize\textwidth\columnwidth\hsize\csname @twocolumnfalse\endcsname
   
\title{
Ising model in small-world networks}
\author{Carlos P. Herrero}
\address{Instituto de Ciencia de Materiales,
         Consejo Superior de Investigaciones Cient\'{\i}ficas (C.S.I.C.),
         Campus de Cantoblanco, 28049 Madrid, Spain  } 
\date{\today}
\maketitle

\begin{abstract}
The Ising model in small-world networks generated from two- and
three-dimensional regular lattices has been studied.
Monte Carlo simulations were carried out to characterize the
ferromagnetic transition appearing in these systems.
In the thermodynamic limit, the phase transition has a mean-field 
character for any finite value of the rewiring probability $p$,
which measures the disorder strength of a given network.
For small values of $p$, both the transition temperature and
critical energy change with $p$ as a power law. 
In the limit $p \to 0$, the heat capacity at the transition 
temperature diverges logarithmically in two-dimensional (2D) networks
and as a power law in 3D.
 \\
\end{abstract}

%

\vskip2pc]
\narrowtext            

\section{Introduction}

Complex networks describe many systems in nature and society,
and have been modeled traditionally by random graphs \cite{bo85}.
In the last years, new models of complex networks have been
introduced, motivated by empirical data in different fields \cite{st01}. 
In particular, small-world networks have been studied, as
they are suitable to describe properties of physical systems with
underlying networks ranging from regular lattices to random graphs,
by changing a single parameter \cite{wa99}.
Watts and Strogatz\cite{wa98} proposed a model for this kind of
networks, which is based on a locally highly connected regular lattice,
in which a fraction $p$ of the links between nearest-neighbor sites
are randomly replaced by new random links, thus creating long-range
``shortcuts''.
The networks so generated are suitable to study different kinds of 
physical systems, such as neural networks \cite{la00} and man-made
communication and transportation systems \cite{wa98,la01,ne00,al99}.

These small-world networks interpolate between the two
limiting cases of regular lattices ($p=0$) and random graphs
($p=1$). In the small-world regime, a local neighborhood is preserved
(as for regular lattices), and at the same time some global properties
of random graphs are maintained.
The small-world effect is usually measured
by the scaling behavior of the characteristic path length $\ell$, 
defined as the average of the shortest distance between two sites.
 For a random network one has a logarithmic increase of $\ell$ with 
the network size $N$ (i.e., the number of sites),
while for a $d$-dimensional regular lattice one expects an algebraic
increase: $\ell \sim N^{1/d}$.    In contrast,
for a small-world network, $\ell$ follows the scaling
law\cite{ba99b,ba00,sc01}
\begin{equation}
\ell(N,p) \sim (N^*)^{1/d} F( N / N^*)   \,,
\label{ell}
\end{equation}
where the scaling function $F(u)$ has the limits $F(u) \sim u^{1/d}$
for $u\ll 1$ and $F(u) \sim \ln u$ for $u\gg 1$; $N^* \sim p^{-1}$ is
a crossover size that separates the large- and small-world regimes
\cite{ba99b,ba99a,me00}.
This indicates that the small-world behavior appears for any finite
value of $p$ ($0<p<1$) as soon as the network is large enough, and in particular
the global characteristics of the network are changed dramatically 
in the presence of only a small fraction of shortcuts.
      
The importance of this shorter global length scale
has been studied for several statistical physical problems
on small-world networks.
Among these problems, one finds in the literature the signal propagation
speed~\cite{wa98}, the spread of infections \cite{ku01,mo00a}, 
as well as site and bond percolation \cite{mo00a,ne99,mo00b}.
The localization-delocalization transition of electron states
has been also studied on quantum small-world networks \cite{zh00}.
                    
Up to now, most of the published work on small worlds has concentrated 
on networks obtained from one-dimensional lattices.
Barrat and Weight\cite{ba00} and Gitterman\cite{gi00} have studied
the crossover from 1D to mean-field behavior for the ferromagnetic
Ising model, which presents a phase transition of mean-field type for
any value of the rewiring probability $p > 0$, provided that the system
size is large enough. 
Close to $p = 0$ the transition temperature $T_c$ goes to zero as
$1/|\log p|$.
A mean-field-type behavior has been also found for the XY model in
small-world networks generated from one-dimensional chains \cite{ki01}.

More recently, Svenson and Johnston\cite{sv01} have studied the damage
spreading for Ising models on small-world networks obtained by rewiring
two-dimensional (2D) and three-dimensional (3D) regular lattices.
They found that these networks are more suitable than regular lattices 
to study social systems with the Ising model.

Here we investigate the ferromagnetic transition for the Ising model 
in small-world networks generated by rewiring 2D and 3D lattices. Contrary to
the networks generated from 1D lattices, now
one has phase transitions at finite temperatures $T_c > 0$ for $p = 0$.
This means that one expects a change from an Ising-type transition
at $p = 0$ to a mean-field-type one in the small-world regime.
We employ Monte Carlo (MC) simulations to obtain average magnitudes 
for finite-size systems. The resulting quantities are extrapolated
to the thermodynamic (infinite size) limit, where the
small-world behavior is expected to dominate the thermodynamic properties
for any finite value of the rewiring probability $p > 0$.

In Sec.\,II we describe the computational method. 
In Sec.\,III we present results of the MC simulations along
with a discussion.
The paper closes with some concluding remarks in
Sec.\,IV.

\section{Computational method}

We consider the Hamiltonian:
\begin{equation}
H = - \sum_{i < j} J_{ij} S_i S_j   \, ,
\end{equation}
where $S_i = \pm 1$ ($i = 1, ..., N$), and the coupling matrix
$J_{ij}$ is given by
\begin{equation} 
\label{Jij}
J_{ij}  \equiv \left\{
     \begin{array}{ll}
         J (> 0), & \mbox{if $i$ and $j$ are connected,} \\
         0, & \mbox{otherwise.}
     \end{array}
\right.
\end{equation}    

\begin{figure} 
\vspace*{-1.5cm}  
\centerline{\psfig{figure=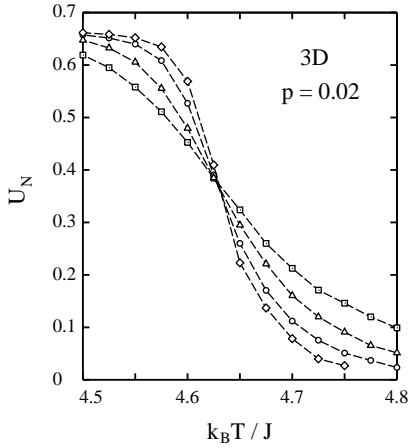,height=9.0cm}}
\vspace*{-1.5cm}  
\caption{
Fourth-order Binder's cumulant $U_N$ as a function of temperature for small-world
networks generated from 3D cubic lattices with rewiring probability
$p = 0.02$. Symbols represent different system sizes $N = L^3$: squares,
$L = 15$; triangles, $L = 20$; circles, $L =25$; diamonds, $L = 30$. 
} \label{f1} \end{figure}

Monte Carlo simulations have been carried out on networks of different
sizes, generated from the 2D square and 3D cubic lattices. 
Small-world networks were generated by randomly replacing a fraction
$p$ of the links of the regular lattices with new random connections.
This procedure keeps constant the total number of links in the
rewired networks. Thus, the average coordination number 
$z$ in the 2D and 3D cases amounts to 4 and 6,
respectively.  In the rewiring process we avoided isolated sites (with zero
links).  With this procedure we obtained networks in which
more than 99.9\% of the sites were connected in a single component
(Note that a random graph has usually many components of various sizes). 
The remaining sites (when they appeared)
were excluded from the final networks employed for the MC simulations.
The size of the networks used in our calculations was larger than the
crossover size $N^*$ \cite{me00,ne99}, so that we were in the small-world 
regime.

The largest networks employed here included 200 $\times$ 200 
sites for the 2D system 
and $40 \times 40 \times 40$ sites for the 3D network. Periodic boundary
conditions were assumed. 
Sampling of the configuration space has been carried out by the Metropolis
local update algorithm \cite{bi97}.  Several thermodynamic quantities
and moments of the order parameter have been calculated for different
simulation-cell sizes.
Finite-size scaling was then employed to obtain the magnitudes corresponding
to the thermodynamic limit (extrapolation to infinite size).
In the remainder of the paper, the presented values for the different
quantities will correspond to the extrapolated ones, unless explicit
mention is made indicating a particular network size.

The ferromagnetic transition temperature has been determined by using Binder's
fourth-order cumulant\cite{bi97}
\begin{equation} \label{Binder}
U_N(T) \equiv 1 - \frac{ \langle M^4 \rangle_N } { 3 \langle M^2 \rangle^2_N }
   \,  ,
\end{equation} 
where the magnetization $M$ of a given spin configuration is given by 
$M = \sum_{i=1}^N S_i / N$.
As an example, we present in Fig. 1 the cumulant $U_N$ as a function of
temperature for different system sizes in the 3D case for $p$ = 0.02.
 The transition temperature
is obtained from the unique crossing point for the different sizes $N$.

\begin{figure} 
\vspace*{-1.5cm}  
\centerline{\psfig{figure=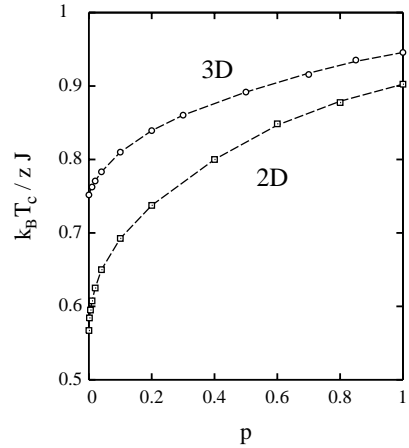,height=9.0cm}}
\vspace*{-1.5cm}  
\caption{
Transition temperature $T_c$, normalized by the average coordination number
$z$, as a function of the rewiring probability $p$ for 
small-world networks generated from 2D and 3D lattices.
Dashed lines are guides to the eye.
} \label{f2} \end{figure}

The heat capacity per site, $c_v$, was obtained from the energy fluctuations
at a given temperature, by using the expression
\begin{equation} 
c_v = \frac {(\Delta E)^2} {N k_B T^2}  \,  ,
\end{equation} 
where $(\Delta E)^2 = \langle E^2 \rangle - \langle E \rangle^2$.

\section{Results and discussion}

In Fig. 2, we present the transition temperature $T_c$ as
a function of the rewiring probability $p$ for 2D and 3D networks.
Similar to the 1D case \cite{ba00}, $T_c$ changes fast close
to $p = 0$, and the derivative $d T_c / d p$ becomes smaller as 
$p$ increases. 
However, the sharp change of $T_c$ for small $p$ shows a behavior
different than that found in the 1D case, where $T_c \sim |\log p|^{-1}$.
In the limit $p = 1$ one finds an increase in $T_c$ as $z$
rises from 4 (in 2D) to 6 (in 3D), as expected for random lattices,
 for which one has $k_B T_c / z J \to 1$ as $z \to \infty$.

\begin{figure} 
\vspace*{-1.5cm}  
\centerline{\psfig{figure=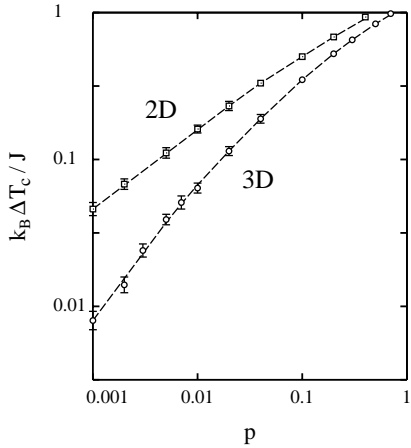,height=9.0cm}}
\vspace*{-1.5cm}  
\caption{ 
Dependence upon the rewiring probability $p$ of the shift in transition
temperature $\Delta T_c$ with respect to the regular lattices, for 2D 
and 3D networks.  Lines are guides to the eye.
} \label{f3} \end{figure} 

To analyze the change in critical temperature with $p$,
we call $\Delta T_c = T_c - T_c^0$, being $T_c^0$ the transition
temperature for the corresponding 2D or 3D regular lattice.
In Fig. 3 we show the dependence of $\Delta T_c$
upon $p$ for the 2D and 3D cases in a log-log plot. 
In both cases we find that $\Delta T_c$ can be fitted by a power law
$\Delta T_c \sim p^s$ for $p \lesssim 0.01$. The exponent $s$
is $0.52 \pm 0.03$ for 2D and $0.96 \pm 0.04$ for 3D.
Our result for 2D networks is compatible with a $\sqrt{p}$
dependence for $\Delta T_c$ near $p = 0$, which means that
the derivative $d T_c / d p$ diverges as 
$\sim 1 / \sqrt{p}$ for $p \to 0$. 
In the 3D case, our results seem to indicate a dependence 
$\Delta T_c \sim p$ for small $p$. However, in this case the
lowest $p$ values studied here may still be too high to attain
the small-$p$ regime (see below).

Associated with the increase in $T_c$ as the rewiring 
probability $p$ rises, one expects an increase in the  
critical energy $E(T_c)$. 
We call $e_c = E(T_c) / N$ the critical energy per site,
and $\Delta e_c = e_c - e_c^0$ its change with respect
to the regular lattice ($p = 0$).
This difference $\Delta e_c$ is shown in Fig. 4 as a function
of $p$ for 2D (squares) and 3D (circles) networks, in a log-log
plot.
For $p \lesssim 0.01$, $\Delta e_c$ can be well fitted in both 
cases by a power law of the form $\Delta e_c \sim p^u$.
For the exponent $u$, we find $u = 0.43 \pm 0.03$ and $0.56 \pm 0.04$
in 2D and 3D, respectively.

\begin{figure} 
\vspace*{-1.5cm}  
\centerline{\psfig{figure=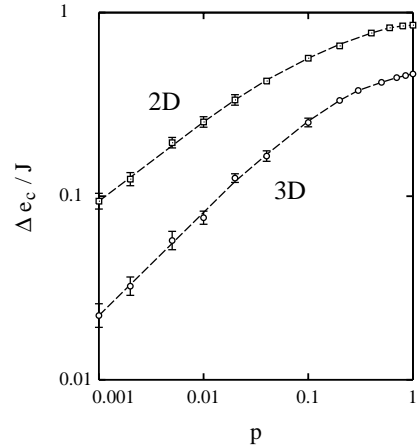,height=9.0cm}}
\vspace*{-1.5cm}  
\caption{
Dependence on the probability $p$ of the shift in critical energy per site
$\Delta e_c$ with respect to the regular lattices, for 2D and 3D networks.
Lines are guides to the eye.
} \label{f4} \end{figure}

A characterization of the ferromagnetic phase transition in these
networks requires the determination of the universality class to which
it corresponds. In the limit $p = 0$ (regular lattices),
one has transitions of the 2D and 3D Ising type. 
In order to determine the type of the phase transition at $p > 0$, we have
studied the critical exponent $\beta$, which gives the temperature 
dependence of the order parameter close to the transition temperature:
$\langle M \rangle \sim (T_c - T)^{\beta}$ for $T < T_c$. 
For the different values of the rewiring probability $p$ studied
here, we have calculated numerically the logarithmic derivative 
\begin{equation}
\mu(t) =  \frac{d \log \langle M \rangle} {d \log t} 
\end{equation}
for $t = T_c - T > 0$, which is related to the exponent $\beta$
through the limit $\beta = \lim_{t \to 0} \mu(t)$.

In Fig. 5 we present results for the derivative $\mu$ 
as a function of temperature for several values of $p$
and for a 2D network of size $200 \times 200$.
For reference, we also present results of MC simulations for $p = 0$
(Ising model on a regular 2D lattice) for the same system size, which
converge to $\beta = 0.125$, the critical exponent for the 2D Ising model.
In all cases $p > 0$, the extrapolation $T \to T_c$ gives an exponent $\beta$
close to 0.5, the value corresponding to a mean-field-type transition.
However, for decreasing $p$, the observation of the mean-field character
of the transition requires to go to temperatures closer to $T_c$ (or 
equivalently to larger system sizes; see \cite{bi97}).
 Thus, for $p = 0.001$ we are still
far from the value 0.5 at the temperatures at which the employed system
size allows us to give a precise value for the order parameter 
$\langle M \rangle$. But even in this case, the departure from the
value expected for a 2D Ising-type transition is clear close to $T_c$.

\begin{figure} 
\vspace*{-1.5cm}  
\centerline{\psfig{figure=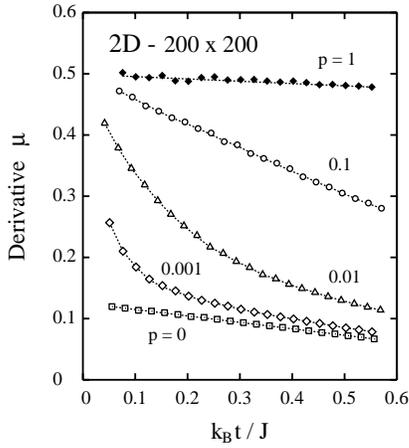,height=9.0cm}}
\vspace*{-1.5cm}  
\caption{
Logarithmic derivative $\mu$ versus the temperature difference $t = T_c - T$ for
small-world networks generated by rewiring a 2D lattice of size $200 \times 200$.
Different symbols represent results obtained for several values of the
rewiring probability $p$. From top to bottom: $p$ = 1, 0.1, 0.01, 0.001, and 0.
} \label{f5} \end{figure}

A complementary way to confirm the mean-field character of the
phase transition consists in determining the exponent $\nu$, which controls
the behavior of the correlation length near the critical temperature:
\begin{equation} \label{xi}
\xi \sim |T - T_c|^{-\nu}  \,  .
\end{equation}    
Close to $T_c$, this critical exponent is related to the temperature
dependence of the fourth-order cumulant defined in Eq. (\ref{Binder}) as
\cite{ki01,bi97}:
\begin{equation} \label{eq:Binder2}
U_N(T) \approx U^* + U_1\left(1-\frac{T}{T_c}\right)N^{1/{\nu}} \, ,
\end{equation} 
with $U^*$ and $U_1$ independent of $T$ and the system size $N$.
From this expression we have:
\begin{equation} \label{eq:Binder3}
\frac{\Delta U_N} {\Delta T} \propto - N^{1/\nu},
\end{equation} 
which allows us to calculate the exponent $\nu$ from the cumulant
$U_N$ derived from the MC simulations.
The values of $\nu$ obtained by this method agreed in all cases 
(within error bars)
with the critical exponent corresponding to mean-field
transitions: $\nu = 0.5$.

We conclude that the mean-field character of the phase transition (which is
the one found in random networks) should
appear in the thermodynamic limit for any $p > 0$. This is in line with
the observation mentioned above that the characteristics of the random graphs
(e.g., mean distance between sites $l \sim \log N$) show up in small-world
networks as soon as their size $N$ is large enough. 

To understand the dependence on $p$ of the transition temperature $T_c$
for small $p$, we will consider the two length scales present in this
problem. On one side we have the correlation length $\xi$, which near 
$T_c$ follows the temperature dependence given in Eq. (\ref{xi}). 
On the other side, we have a length scale characteristic of the small-world
network, given by the typical distance between ends of shortcuts:
$\zeta = (p z)^{-1/d}$ \cite{ne99}.
When the correlation length is smaller than $\zeta$, the system behaves
basically as a regular lattice. When $\xi$ grows beyond $\zeta$
(as happens when we approach the transition temperature from above,
$T \to T_c^+$), the ``long-range'' interactions introduced by the
shortcuts come into play and give rise to the mean-field behavior.
Thus, the transition between the regular-lattice behavior and the 
mean-field one occurs for $\xi \approx \zeta$.
For the 1D Ising model, the correlation length at low temperatures
($k_B T \ll J$) is given by $\xi \sim \exp(2 J / k_B T)$ \cite{bi92}. 
Taking into account that in this case 
$\zeta \sim 1/p$, the condition $\xi \approx \zeta$ suggests a 
critical temperature for the 1D small-world $T_c \sim |\log p|^{-1}$,
in agreement with more detailed calculations for this system \cite{ba00}.
 For the 2D Ising model the correlation length near $T_c$ scales as 
$\xi \sim |T - T_c^0|^{-1}$ \cite{mc73,bi92}, whereas now 
$\zeta \sim 1/p^{1/2}$.
Then, using the same argument, one expects for small-world 
networks generated from 2D lattices:
$T_c - T_c^0 \sim p^{1/2}$, in good agreement with the results of
our Monte Carlo simulations (see Fig. 3). Thus, it seems that in general,
for systems with a correlation length diverging as $|T - T_c^0|^{- \nu}$,
the order-disorder transition temperature for small $p$ depends on the rewiring
probability as 
\begin{equation} \label{Tc}
T_c - T_c^0 \sim p^{1/\nu d}   \, .
\end{equation}
Note that here $\nu$ is the critical exponent of the considered model in the 
regular lattice, not the one corresponding to the phase transition in the
small-world network (which is the mean-field one).
 Hence, we see that the condition $\xi \approx \zeta$ describes the transition
between the large-world regime (which corresponds to regular lattices)
and the small-world one \cite{me00}, as well as the ferromagnetic transition,
indicating that the small-world transition and the order-disorder one are
intimately related.

\begin{figure} 
\vspace*{-1.5cm}  
\centerline{\psfig{figure=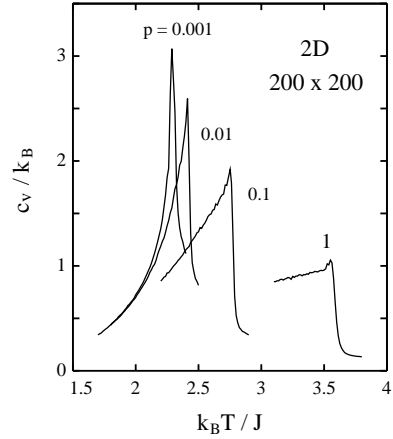,height=9.0cm}}
\vspace*{-1.5cm}  
\caption{
Heat capacity per site $c_v$ versus temperature for small-world 
networks generated from a 2D lattice of size $200 \times 200$.
The plotted curves correspond to different values of the rewiring
probability $p$, as indicated by the labels.
} \label{f6} \end{figure} 

The above argument, however, seems to fail for small-world networks
built up from 3D lattices. For the 3D Ising model one has a critical
exponent $\nu \approx 0.63$ \cite{bi92}, which should give for small $p$ a
dependence: $\Delta T_c  \sim p^{0.53}$, with an exponent clearly
smaller than that derived from our MC simulations, where we found
for $p \lesssim 0.01$: $\Delta T_c  \sim p^{0.96}$.
It seems that this discrepancy appears because the smallest $p$ value
employed in our simulations ($p = 10^{-3}$) is still too large to 
observe the small-$p$ behavior. Although for such values of $p$ we are
clearly in the small-world regime, the corresponding value of $\zeta$
in 3D ($\zeta \sim$ 5), is too small to allow the correlation length
to reach the critical dependence of the Ising model in the regular lattice:
 $\xi \sim |T - T_c^0|^{-\nu}$.
In other words, to observe the $p$-dependence of $T_c$ given in Eq. (\ref{Tc}),
one needs $\zeta$ values larger (i.e., smaller $p$ values or larger networks)
than those employed here.
A similar conclusion was proposed for the XY model on 1D small-world
networks \cite{ki01}, for which MC simulations gave a dependence
$T_c \approx a \log p + b$ (with $a$ and $b$ numerical constants)
instead of $T_c \sim p$, expected from the above argument.

For mean-field-type transitions, the heat capacity per site $c_v$ 
shows a finite jump (not a divergence) at $T_c$.
Given that at $p = 0$ one has Ising-type phase transitions with divergent
heat capacity at $T_c$, such a jump in $c_v$ will diverge in the limit $p \to 0$.
The temperature dependence of $c_v$ is displayed in Fig. 6 for several
values of $p$ and for networks built up from a $200 \times 200$ 2D lattice. 
As expected, one observes an increase in the maximum value of $c_v$
as $p$ is reduced.

\begin{figure}
\vspace*{-1.5cm}
\centerline{\psfig{figure=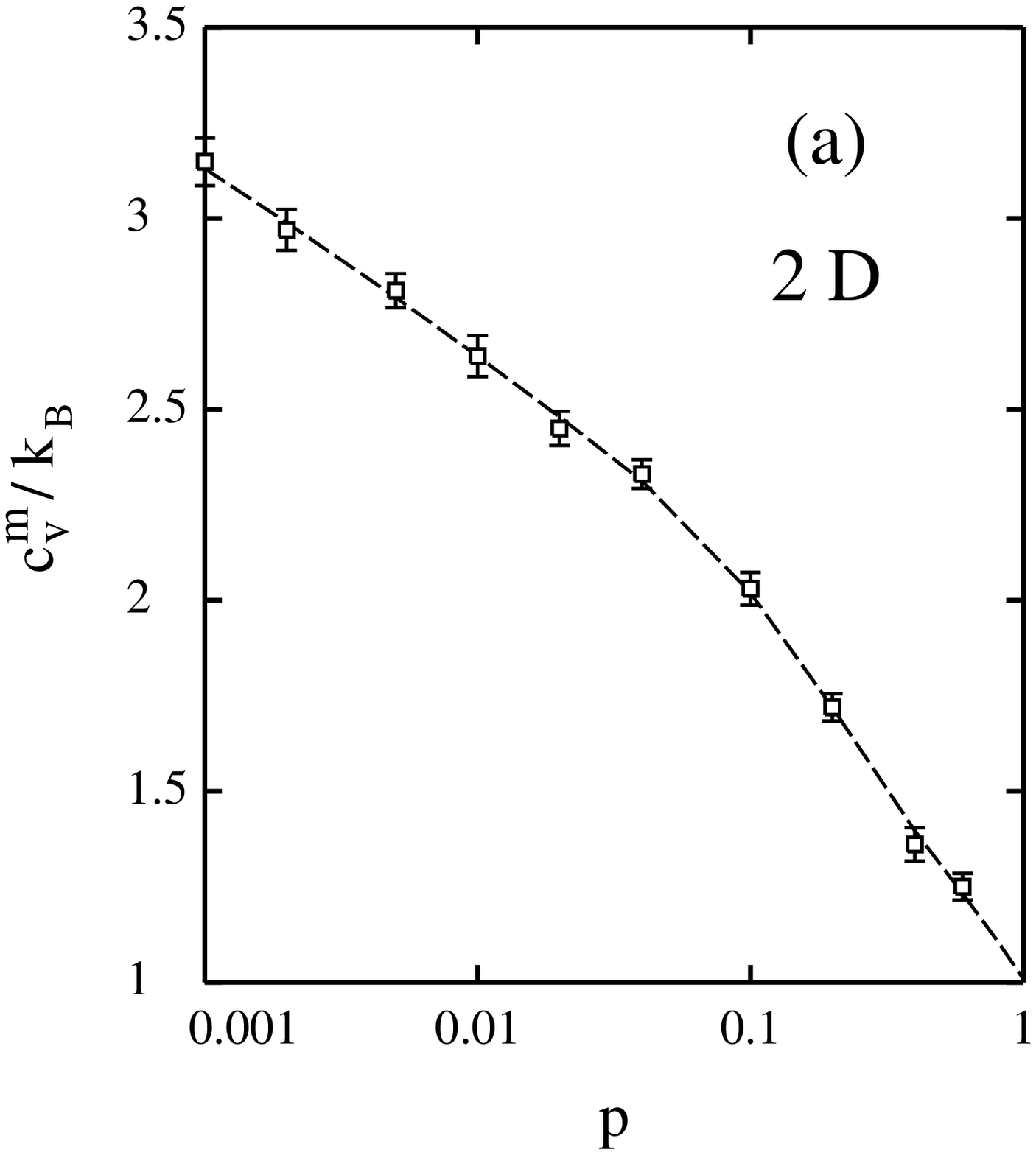,height=9.0cm}}
\vspace*{-2.5cm}
 
\centerline{\psfig{figure=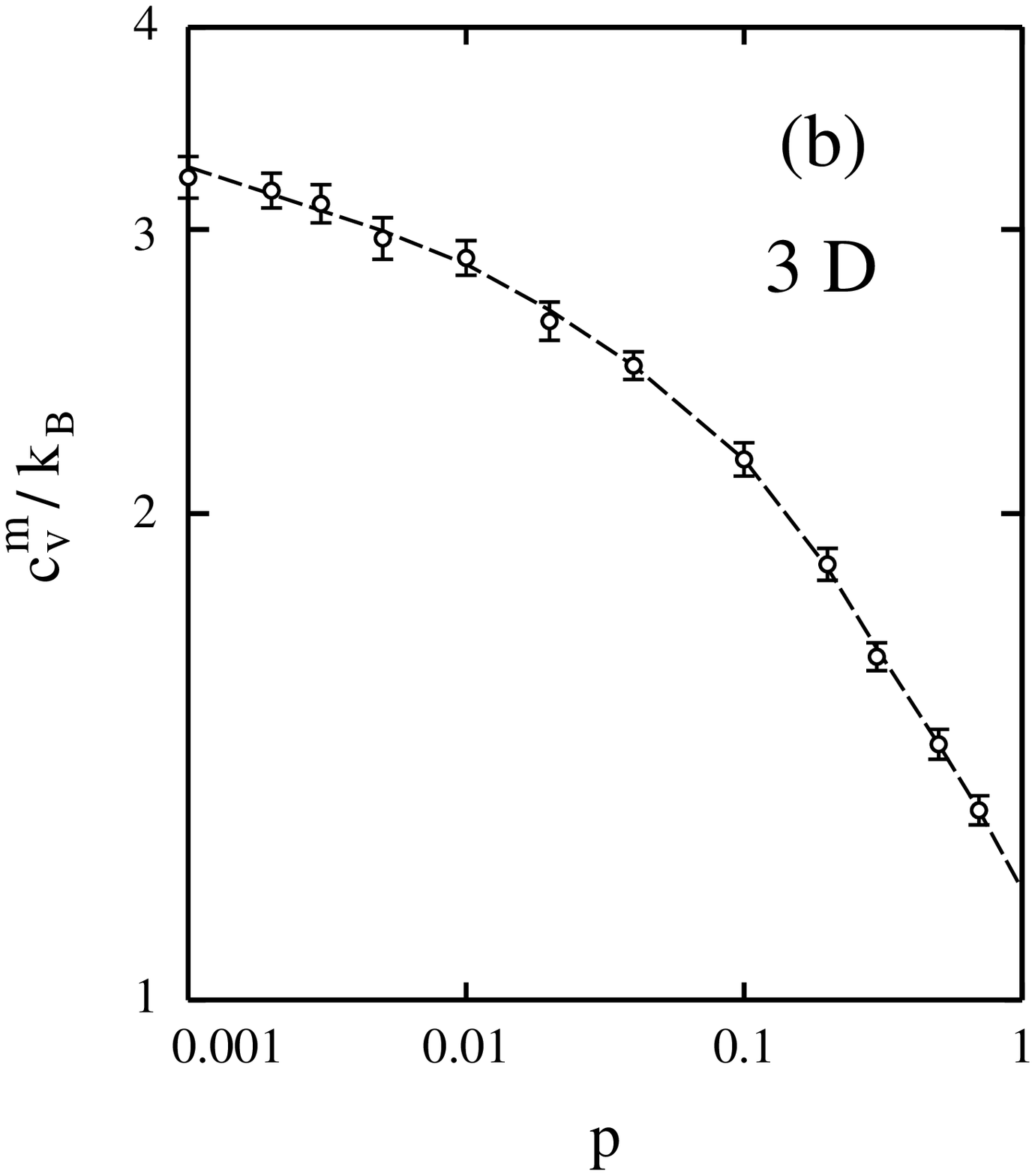,height=9.0cm}}
\vspace*{-1.2cm}
\caption{
Maximum value of the heat capacity per site $c_v^m$ as a function of the
 rewiring probability $p$.
Symbols correspond to the extrapolation of finite-size results to
$N \to \infty$.
(a) 2D networks, in a semi-logarithmic plot; (b) 3D networks, in a
log-log plot.  Dashed lines are guides to the eye.
} \label{f7} \end{figure}
         
The maximum value of $c_v$ for each $p$, extrapolated to the thermodynamic
limit, is shown in Fig. 7 for 2D and 3D small-world networks.
In Fig. 7(a) we present our results for 2D, which
for small $p$ display a logarithmic dependence $c_v^m = a - b \ln p$, with
the numerical constants $a = 1.74 \pm 0.06$ and $b = 0.20 \pm 0.01$.
This logarithmic dependence of the heat capacity can be explained by
arguments similar to those employed above to analyze the behavior of
the transition temperature as $p \to 0$.
For the Ising model in 2D regular lattices, the heat capacity near
$T_c^0$ diverges as:
 $c_v(T) \sim - \log |T -T_c^0|$ \cite{mc73}.
Taking into account the relation between $T_c$ and $p$ given above for
small $p$, one finds for the 2D small-world networks
$c_v^m \sim - \log p$, in agreement with the results of our Monte Carlo
simulations.

In the 3D Ising model, $c_v$ has at $T_c^0$ a singularity of the form
$|T -T_c^0|^{-\alpha}$, with $\alpha \approx 0.12$ \cite{bi92}. Assuming that
$T_c - T_c^0 \sim p^{1/\nu d}$, as in Eq. (\ref{Tc}), one expects for
$c_v^m$ close to $p = 0$ a power-law of the form: $c_v^m \sim p^w$, with
an exponent $w = - \alpha / \nu d = - 0.063$.
In Fig. 7(b) we present the values of $c_v^m$ derived from our
MC simulations as a function of $p$ in a log-log plot.
These results are consistent with a power-law dependence
$c_v^m \sim p^w$ at small $p$. In fact, for $p \lesssim 0.01$
our results can be fitted with an exponent $w = - 0.057 \pm 0.005$.
However, as in the case of the transition temperature discussed above,
smaller values of $p$ are necessary to determine unambigously
this dependence from numerical simulations, and in particular to find
the exponent $w$.

\section{Concluding remarks}
We have studied the ferromagnetic transition that appears for the Ising
model in small-world networks generated from 2D and 3D regular lattices.
In these networks, the presence of a small disorder ($p > 0$)
causes a change in the universality class of the order-disorder 
transition, from Ising for $p = 0$ to mean-field type for $p > 0$.

Our results indicate that the order-disorder transition  
occurs at a temperature $T_c$ where
the spin correlation length $\xi$ is on the order of 
the length $\zeta$ (typical distance between ends of shortcuts),
characteristic of these networks.
In particular, close to $p = 0$ this gives a dependence
$T_c \sim |\log p|^{-1}$ in 1D networks, and
$T_c - T_c^0 \sim p^{1/\nu d}$ for networks generated from
regular lattices of higher dimensions.
This is the dependence found from our Monte Carlo simulations for
2D networks. In the 3D case we find a power law for $T_c - T_c^0$,
but the determination of the actual exponent from MC simulations 
for $p \to 0$ requires $p$ values smaller (i.e., networks
larger)  than those employed here.

From the results presented in this paper, 
it is clear that there is a close relation between the small-world
transition and the order-disorder transition in this kind of networks.

The author benefited from useful discussions with M. A. R. de Cara and 
M. Saboya. Thanks are due to E. Chac\'on for critically reading the
manuscript.  This work was supported by CICYT (Spain) under Contract 
No. PB96-0874, and by DGESIC through Project No. 1FD97-1358.  \\


\end{document}